\documentclass[10pt,a4paper,twocolumn,aps,pre,showkeys,showpacs]{revtex4-1}
\usepackage{graphicx}
\usepackage{dcolumn}
\usepackage{bm}
\usepackage[utf8]{inputenc}     
\usepackage{amsmath,amssymb,amsfonts}
\usepackage{amsthm}
\usepackage{textcomp} 
\usepackage{xcolor}
\usepackage{graphicx} 
\DeclareGraphicsExtensions{.eps,.png,.jpg,.pdf}
\usepackage{float}

\begin{document}


\title{An entropic simulational study of the spin-$1$ Baxter-Wu model in a crystal field}

\author{L. N. Jorge}
\affiliation{Instituto Federal de Educação, Ciência e Tecnologia do Mato Grosso Campus Cáceres - Prof. Olegário Baldo, CEP 78201-380, Cáceres, Mato Grosso. Brazil}
\email{lucas.jorge@cas.ifmt.edu.br}

\author{P. H. L. Martins}
\affiliation{Instituto de Física, Universidade Federal de Mato Grosso, CEP 78060-900, Cuiabá, Mato Grosso, Brazil.}

\author{Claudio J. DaSilva}
\affiliation{Instituto Federal de Educação, Ciência e Tecnologia de Goiás, Rua 76, Centro, Goi{\^a}nia - GO, Brazil.}

\author{L. S. Ferreira}
\affiliation{Instituto de Física, Universidade Federal de Goiás, C.P. 131 CEP 74001-970, Goiânia, Goiás, Brazil.}

\author{A. A. Caparica}
\affiliation{Instituto de Física, Universidade Federal de Goiás, C.P. 131 CEP 74001-970, Goiânia, Goiás, Brazil.}

\begin{abstract}
We investigate the critical behavior of the two-dimensional spin-$1$ Baxter-Wu model in a crystal field using entropic sampling simulations with the joint density of states. We obtain the temperature-crystal field phase diagram, which includes a tetracritical line ending at a pentacritical point. A finite-size scaling analysis of the maximum of the specific heat, while changing the crystal field anisotropy, is used to obtain a precise location of the pentacritical point. Our results give the critical temperature and crystal field as $T_{pc}=0.98030(10)$ and $D_{pc}=1.68288(62)$. We also detect that at the first-order region of the phase diagram, the specific heat exhibits a double peak structure as in the Schottky-like anomaly, which is associated with an order-disorder transition.

\end{abstract}

\maketitle


\section{\label{sec:level1}Introduction}\label{sec:introduction}

The spin-$1$ Baxter-Wu (BW) model in a crystal field\cite{Kinzel1981,Costa2004,Dias2017} is a generalization of the original spin$-\frac{1}{2}$ BW model\cite{Wood1972,Baxter1973,baxter1974ising,baxter1974ising2}, which includes a crystal field anisotropic term $D$, in addition to the three-spin interaction. The Hamiltonian of the model considered here is
\begin{eqnarray}\label{Eq.HBWD}
\mathcal{H}= -J\sum_{\langle i,j,k\rangle}s_{i}s_{j}s_{k}+D\sum^N_i s_i^2, \label{eq.campo}
\end{eqnarray}
where the spin variables are located at the sites of a triangular lattice and assume the values $s_i = \pm 1,0$, $J>0$ is the ferromagnetic coupling
constant that defines the energy scale, and $D$ is the anisotropy due to the crystalline field. The first sum extends over all triangular faces while the second runs over all lattice sites. Note that, for the spin-$\frac{1}{2}$ case, the second term in Eq. \ref{eq.campo} is just a constant.
The Hamiltonian in Eq. \ref{Eq.HBWD} resembles the Blume-Capel (BC) model\cite{Blume1966,Capel1966} case, which is described by a phase diagram that presents a line of continuous phase transitions meeting a discontinuous one at a tricritical point.

In the thermodynamic limit, at zero temperature and for $D < 2$, four ordered states are present in the ground state, one ferromagnetic, $(+++)$, and other three ferrimagnetic phases: $(+--)$, $(--+)$ and $(-+-)$, so that the spin-$1/2$ BW model is recovered. For $D>2$, the state that minimizes the energy is the microstate $(000)$, with all spins $s_i=0$. For $D=2$, one has a first-order phase transition with all those five phases coexisting.  

The first proposal to study the BW model with interaction anisotropy was made by Kinzel \textit{et al.}\cite{Kinzel1981}, analyzing the correlation length behavior, the authors concluded that a continuous transition only occurs for the pure case, $s=1/2$ and for any other crystal field value, there is a discontinuous transition. The BW model with crystal field anisotropy was studied by Costa\cite{Costa2004} using the renormalization group, conventional finite-size scaling, and conformal invariance technique. The existence of a pentacritical point was conjectured at $D_c=1.3089$ and $T_c=1.2225$. Recently, Dias 
\textit{et al.}\cite{Dias2017} by applying finite-size scaling and conformal invariance, analyzed both the spin-$1$ and spin-$3/2$ cases. For the latter, they extended the size of the strips that are used in reference\cite{Costa2004} and found $D_c=0.890254$ and $T_c=1.1690$. A phase diagram similar to the BC model was obtained for both cases.

The case with $D=0$ was studied using entropic simulations, resulting in the observation that the BW model presents a mixture of phase transitions\cite{jorge2020order}, related to a tetracritical point. That was settled after the analysis of the energy probability, the configurations in the critical region, and by separating the density of states in two parts, sorting out the ferromagnetic and ferrimagnetic arrangements. Besides, this previous study analyzed how the choice of lattice sizes affects the results on the thermodynamic limit for the critical exponents and critical temperature, and it is clear now that choosing multiple of three lattice sizes or not is equivalent. 

Entropic simulations\cite{Wang2001,Caparica2012,Caparica2014,ferreira2018} are excellent to study phase transitions and critical phenomena. The results obtained by this technique have revealed important characteristics regarding different models\cite{Caparica2015,jorge2020order,jorge2019three,Ferreira2020}.  The estimation of the joint density of states\cite{Zhou2006,ferreira2018,Ferreira2020} brings a range of additional information, besides making it possible to obtain thermodynamic properties for any temperature and coupling constants, allowing the construction of the phase diagram.

In the present work, we completely characterize the phase diagram and determine the position of the multicritical point by studying the model using the joint density of states. We also analyze the possibility that the line, said to be of second-order, might be a line of coexistence of four configurations ending in a pentacritical point. Finally, we perform a study of the specific heat in the region of the first-order transition to understand what causes this transition.  

This work is organized as follows: in the next section, we describe the computational details. In section \ref{sec:phasediagram}, we present the results and the phase diagram, and we still report an anomaly that we found out in the specific heat. The last section comprises the conclusions and some further remarks.

\section{Entropic sampling simulations with the joint density of states}\label{sec:joint}

The study of the effects of spin anisotropy in a magnetic system has arisen interest in many situations\cite{plascak2003universality,Bahmad2007,Zierenberg2015}. An appropriate approach to tackle these problems is to construct a joint density of states\cite{Kwak2015wang}, with a second parameter that characterizes the model. To investigate dimers, some authors have used a joint density of states with three-variables dependence\cite{interactingdimers2019}, $g(N_3,U,N)$, where $N_3$ is the number of the energetic dimers, $U$ is the energy between interacting dimers, and $N$ is the total number of dimers. 

In the present case, entropic simulations with a joint density of states can be performed by defining 
\begin{equation}
 E_1=\sum_{\langle i,j,k\rangle}s_is_js_k, 
\end{equation}
and
\begin{equation}
  E_2=\sum_is_i^2.
\end{equation}
Hence the total energy assumes the form  $E=-JE_1+DE_2$.
The probability of flipping from a configuration with $ (E_1, E_2) $, to a new one, with $ (E_1 ', E_2') $, is given by

\begin{equation}
 p[(E_1,E_2)\to (E_1',E_2')]=\textmd{min}\left(\frac{g(E_1,E_2)}{g(E_1',E_2')},1\right).
\end{equation}
 
After each Monte Carlo step (MCS), the joint density of states and the histogram are updated as $\ln g(E_1, E_2) \to \ln g(E_1, E_2) + \ln f $, where $f$ is the modification factor, initially set as $f_0=e=2.71828...$ and $H(E_1,E_2) \to H(E_1,E_2)+1$, respectively\cite{Caparica2012}. It is important to mention that each MCS consists of flipping $L^2$ spins, whether the attempt is accepted or not. This procedure continues until the histogram reaches its flatness -- i.e., each term of the histogram is at least $80\%$ of its average over all the possible values of $E_1$ and $E_2$. Then the modification factor is updated to $f_ {i + 1}=\sqrt{f_i}$ and the histogram is reset. Furthermore, to halt the simulations, we use the parameter $\epsilon=|T_c(t)-T_c(0)|$\cite{Caparica2014}. To calculate $\epsilon$, we use the following procedure: beginning from the modification factor $f_{13}$, when the flatness criterion is satisfied, the temperature of the peak of the specific heat $T_c(0)$ is calculated, using the current density of states at the end of each Wang-Landau (WL) level, and whenever the flatness of the histogram is verified, its value is updated. If $\epsilon$ remains less than $10^{-4}$ and the histogram satisfies the flatness condition for the same level of WL, then the microcanonical averages and the density
of states are saved and the simulation is finished. For all lattice sizes, simulations started from a single run until the Wang-Landau level $f_6$, since up to that point the current density of states is not yet biased and can proceed to any final result. This procedure saves about 60\% of CPU time, in comparison with the case of starting from the first Wang-Landau level. Ref. \onlinecite{jorge2019three} applied this approach with great success.

According to the canonical ensemble, the partition function can be written as
\begin{equation}
 Z(T,D)=\sum_{E_1,E_2}g(E_1,E_2)e^{(JE_1-DE_2)/k_BT},
\end{equation}
where $g(E_1,E_2)$ is the joint density of states. So, for each lattice size, we set up $g(E_1,E_2)$ in order to obtain the thermodynamic quantities for any temperature and any value of the crystal field $D$. 

Since the function $g(E_1,E_2)$ does not depend on temperature or the crystal field, we can calculate any thermodynamic quantity  without performing a new simulation run. 

The canonical averages of any thermodynamic quantity $A$ can be calculated as
\begin{equation}\label{mean}
 \langle{A}\rangle_{T,D}=\dfrac{\sum_{E_1,E_2}\langle A\rangle_{E_1,E_2} P(E_1,E_2,J,D)} 
{\sum_{E_1,E_2} P(E_1,E_2,J,D)},
\end{equation}
where $\langle A\rangle_{E_1,E_2}$ is the microcanonical average accumulated during each simulation run and
\begin{equation}
 P(E_1,E_2,J,D)=g(E_1,E_2)e^{(JE_1-DE_2)/k_BT},
\end{equation}
is the canonical distribution. 

Performing entropic simulations with a joint density of states requires huge computational effort, so we considered here only small lattice sizes, yielding excellent results even though.  Ref. \onlinecite{Jorge2016} demonstrated that when using the order parameter as the total magnetization, one can consider systems with non-multiple of three lattice sizes without loss of generality of the model \cite{jorge2020order}. Thus we ran simulations  for $L=8$, $10$, $14$, and $16$ with $n=24$, $20$, $20$, and $16$ independent runs, respectively.

\section{Phase Diagram} \label{sec:phasediagram}
\subsection{Determination of the pentacritical point}

\begin{figure} [!t] 
\vspace{0.1in}
\includegraphics[scale=0.57,angle=-90]{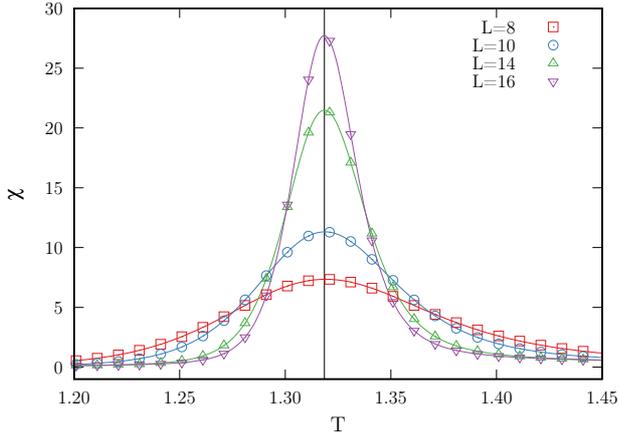}
\vspace{0.10in}
\caption{Magnetic susceptibility for $D=1.100$, for all lattice sizes simulated. The temperature of the peak for each curve falls close to a common value. \label{fig:khi-D}}
\end{figure}

In this work, we determine the critical line of the $D\times T$ phase diagram using the maximum of the magnetic susceptibility for the largest lattice size taken in our simulations, since, as Fig.\ref{fig:khi-D} shows, all maxima are very close to each other, around the same temperature, thereby a finite-size scaling behavior is not visible to the eyes. In fact, finite-size scaling effects for the critical temperature are very subtle when one uses lattice sizes that are not multiples of three\cite{Jorge2016,jorge2020order,jorge2019three}. Thus, the difference between the critical temperature at the thermodynamic limit and those obtained for the finite sizes is small, making the critical line observed in the diagram very close to that expected for an infinite system. Hence, for each $D$ value there is a temperature $T_c$ that corresponds to the maximum of the magnetic susceptibility.

The zero crystal field case, $D=0$, was addressed in references \onlinecite{costa2004monte} and \onlinecite{jorge2020order}. In the first work, the authors performed MC simulations using the Metropolis algorithm and found a critical temperature of $T_c=1.6607(3)$ for a continuous transition. In the second one, the authors carried out an entropic sampling approach and demonstrated that the system displays duplicity in phase transitions, since the scaling laws for continuous and discontinuous transitions are equally applicable, and resulted in practically the same temperature value $T_c=1.660549(51)$ and $T=1.660577(17)$, respectively. The coexistence of continuous and discontinuous phase transitions was discovered by simultaneously generating two densities of states for the ferromagnetic and ferrimagnetic states, which showed the different behaviors of each phase transition. For the ferromagnetic state, there is a discontinuous transition exhibiting a double peak in the energy probability distribution P(E), two inflection points in the inverse of the microcanonical temperature
$\beta=dS/dE$, and a discontinuity in the order parameter around the transition temperature. For the ferrimagnetic ones, a continuous transition, with only one peak in P(E), only one inflection point in $\beta$, and the order parameter with a typical second-order transition behavior. The analysis of the spin configurations showed the coexistence of ferromagnetic, ferrimagnetic, and paramagnetic clusters in different configurations, revealing a tetracritical behavior.

\begin{figure} [!t] 
\vspace{0.1in}
\includegraphics[scale=0.56,angle=-90]{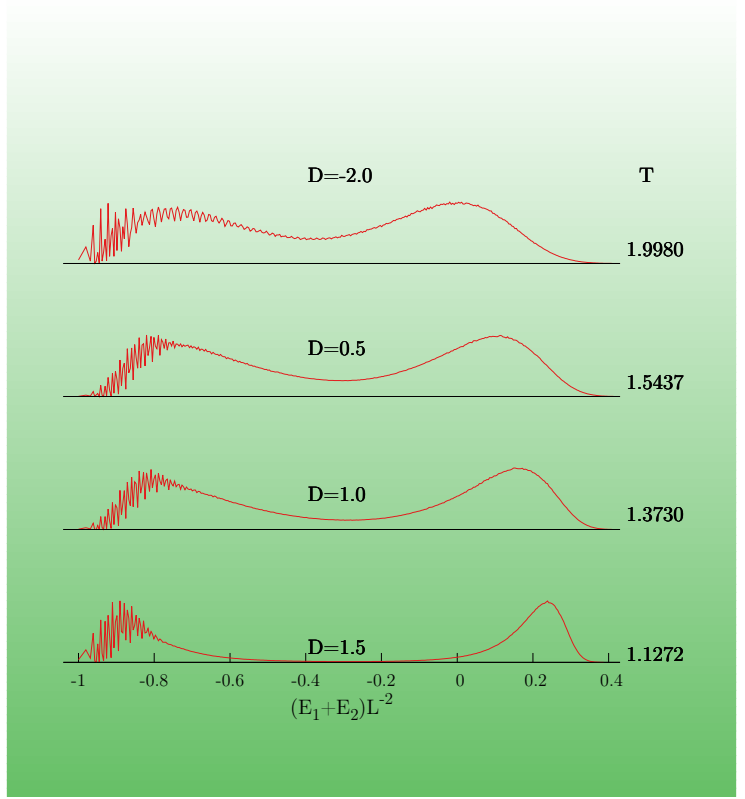}
\vspace{0.3in}
\caption{Energy probability distribution as a function of $(E_1+E_2)/L^2$ for different values of $D$, for $L=16$. The temperatures where the two peaks occur are shown at the right side of the graph.
\label{fig:prob}}
\end{figure}

To find out if the same behavior persists for $D\ne0$, we analyze the energy probability distribution for other values of the crystal field, shown in Fig \ref{fig:prob}. The temperatures at which the two peaks have the same height are shown on the right. The first curve represents the energy probability for $D=-2.0$, where we see two peaks. However, this double peak characteristic can be a finite-size effect and, a more detailed study in this region is still missing. For positive values of the field, the finite-size effect will not remove the two peaks, since we have for the case $D=0.0$ the presence of these two peaks at the thermodynamic limit \cite{jorge2020order} and one can see that the greater the crystal field, the larger the separation between the two peaks. Therefore, we might conclude that the second-order critical line is a line of tetracritical points.

\begin{figure} [!t] 
\includegraphics[scale=0.57,angle=-90]{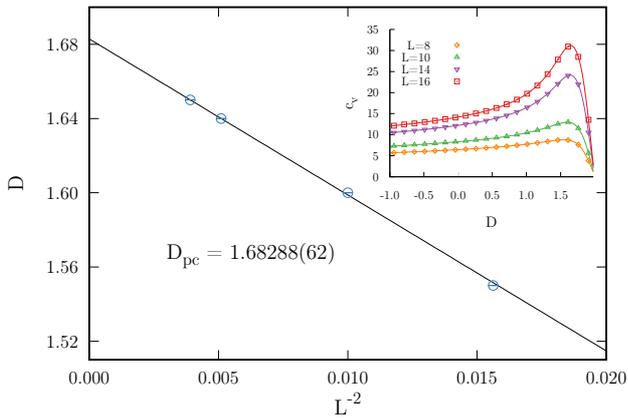}
\caption{Size dependence of the locations of crystal field function at
the extrema in the specific heat for lattice sizes $L=8$, $10$, $14$, and $16$. 
The inset shows behavior of the maximum height of the specific heat while increasing the crystal field. Its maximum value is achieved
at the transition to the paramagnetic phase.\label{fig:fit-Dc}}
\end{figure}

To determine the critical crystal field, we have used the technique described in \onlinecite{Care1993}, where it is calculated from behavior the specific heat maximum, $C_{v_{max}}(D)$. The evolution of the specific heat maximum versus $D$ results in a peak, which is associated to the beginning of a first order phase transition. This behavior is presented in the inset of Fig \ref{fig:fit-Dc}. The critical crystal field is obtained from the scaling law\cite{Zierenberg2015}
\begin{equation}
 D_c(L)=D_c^{\infty}+b L^{-d}.
\end{equation}
Figure \ref{fig:fit-Dc} shows the best fit for $d=2$, which gives $D_c=1.68288(62)$ at the thermodynamic limit. 

At the pentacritical point, it is expected that the maxima of the specific heat and the susceptibility scale with the dimensionality\cite{Fisher1982,yamagata1993first}, e.g.,
\begin{equation}
 T_c(L)=T_c^{\infty}+a_2L^{-d}, \label{eq:fit-T-1a}
\end{equation}
where $T_c^{\infty}$ is the critical temperature at the thermodynamic limit and $a_2$ is a quantity-dependent constant, enabling the determination of $T_c$
by extrapolating $L\to\infty$ $(L^{-d}= 0)$ from the linear fitting given by the temperature of the maxima of the specific heat and the susceptibility.

\begin{figure} [!t] 
\vspace{0.1in}
\includegraphics[scale=0.56,angle=-90]{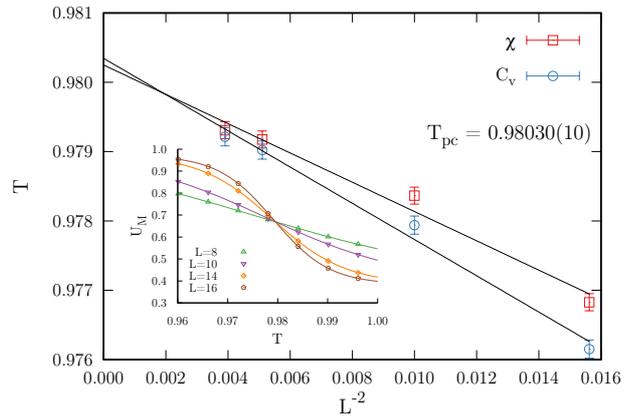}
\vspace{0.10in}
\caption{Temperature of the maxima of specific heat and susceptibility versus $L^{-2}$. The error bars are smaller than symbols. The inset shows the cumulant of the magnetization for the simulated lattice sizes. The crossing of the cumulants takes place in a region very close to that estimated for the pentacritical point.
} \label{fig4}
\end{figure}

Figure \ref{fig4} shows the linear fitting of the temperature of the maxima of specific heat and susceptibility against $L^{-2}$, for the critical crystal field, which converges to $T_c$ when $L^{- 2}\to0$. The value for the temperature of the pentacritical point was $ T_c = 0.98030 (10) $. We reinforce that the critical temperature variation is in the fourth decimal place, as predicted by the susceptibility curves. In the inset we present the cumulants of the  magnetization for the simulated lattice sizes. The crossing of the cumulants takes place in a region very close to that estimated for the pentacritical point. 

\begin{figure} [htb] 
\vspace{0.1in}
\includegraphics[scale=0.56,angle=0]{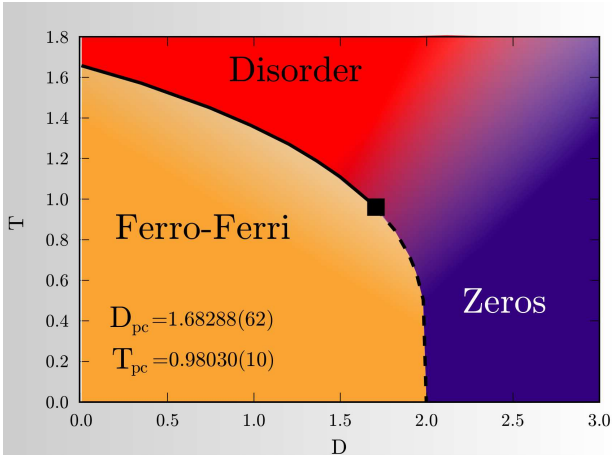}
\vspace{0.10in}
\caption{Phase diagram for the Baxter-Wu spin-$1$ model with crystal field anisotropy in the $D$ versus $T$ plane. The pentacritical point separates the first order phase transition region and criticality.
 \label{fig:TcxD-khi}}
\end{figure}

In Fig. \ref {fig:TcxD-khi}, we show the phase diagram. The continuous line represents the tetracritical points that separate the ferro-ferri phases from disorder. The dashed line  represents the points of coexistence of the five phases: the four fundamental states and the $(0~0~0)$. The point $D=2$ is a fivefold one, where the five states are equally probable. For $D>2$ only the state with spins $s_i=0$ remains, since it corresponds to the minimum of the free energy. The location of the pentacritical point is described in Table \ref{tab:campo}, together with some results from previous works, for comparison. We have obtained $D_c=1.68288(62)$ and $T_c=0.98030(10)$.

\begin{table}[!t] 
\vspace{-0.15cm}
\setlength{\arrayrulewidth}{2\arrayrulewidth}  
   \setlength{\belowcaptionskip}{9pt} 
\begin{tabular}{ccc}

                                                &   $D_c$     &  $T_c$       \\ \hline
                                               Costa\cite{Costa2004} & $1.3089$   & $1.2225$   \\
                                               Dias\cite{Dias2017} & $0.890254$ & $1.1690$ \\
                                               Our results& $1.68288(62)$ & $0.98030(10)$ \\
                                               
     \hline
\end{tabular}
\caption{Comparison of our results for the pentacritical point to those obtained by Costa\cite{Costa2004} and Dias\cite{Dias2017}.}\label{tab:campo}
\end{table}

 \begin{figure} [htb] 
\vspace{0.1in}
\includegraphics[scale=0.56,angle=-90]{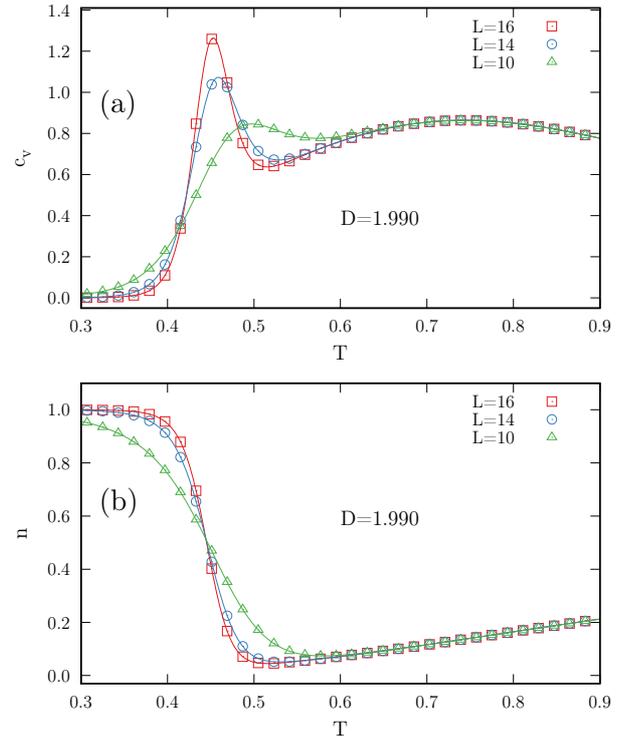}
\vspace{0.10in}
\caption{(a) Double peak structure observed for specific heat at $D=1.990$. The peak on the left is due to the first order phase transition, and on the right is the anomalous peak. (b) Density of spins other than zero at $D=1.990$. It starts at one, and falls dramatically around $T=0.45$.
 \label{fig:cv-schot1}}
\end{figure} 
The discrepancy among the results obtained by different methods is because the continuous phase transition line is, in fact, a line of tetracritical points, where we have the presence of two types of phase transitions, namely, continuous and discontinuous. Therefore, it is acceptable that the previous results have found only one of these points, and not at the pentacritical point. This is also true due to computational difficulties since small strips in the renormalization group technique were used, which leaves the results far from those obtained for the thermodynamic limit. The authors mentioned that regarding the first-order transition line, one can still argue that there is not enough data (because they had to increase the width of the strips by $3$ to accommodate all three different sublattices) to support its characteristic \cite{Dias2017}.
This new interpretation of the line of continuous transitions has been complemented previous works and allows authors to review their results. To the best of our knowledge, this duality of the BW model was unknown until now.

\subsection{Anomaly of the specific heat for large crystal field in the first order region
}\label{subsec:anomalies}

By analyzing the specific heat, we detected a unusual behavior for large values of the crystal field, above $D=1.900$. An anomaly appears besides its maximum, and it does not seem to scale with the size of the system. Above this field value, a small peak in specific heat appears around $T=0.80$, but it becomes more evident as the value of the crystal field increases (in fact, this second peak does not increase, but the first peak decreases as one approaches $D=2$).

\begin{figure} [!t] 
\vspace{0.1in}
\includegraphics[scale=0.56,angle=-90]{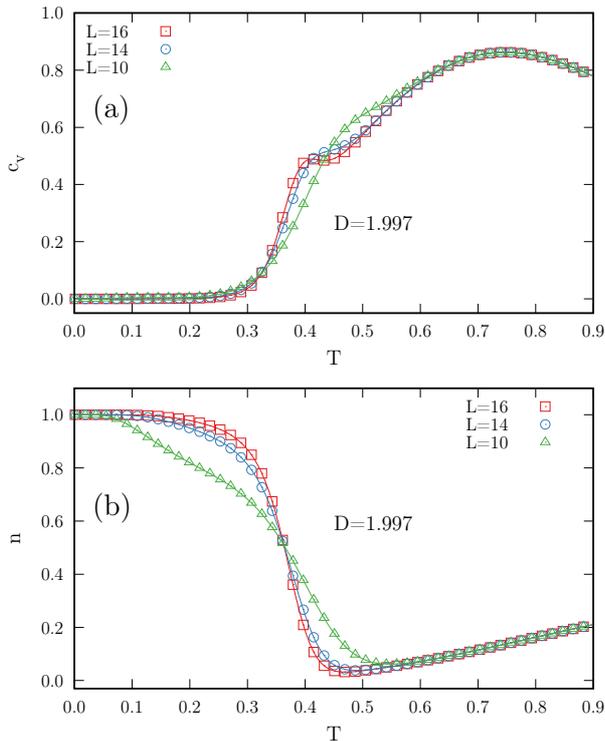}
\vspace{0.10in}
\caption{(a) Double peak structure observed for specific heat at $D=1.997$. The peak on the left is due to the first order phase transition, and on the right is the anomalous peak. (b) Density of spins other than zero and equal to zero at $D=1.997$. It starts at one and fall dramatically around $T=0.40$. \label{fig:cv-schot3}}
\end{figure}

In Figs. \ref{fig:cv-schot1}a and \ref{fig:cv-schot1}b, the anomaly in $C_v$ and the density of spins are shown for $D=1.990$. Figures \ref{fig:cv-schot3}a and \ref{fig:cv-schot3}b show the anomaly in $C_v$ and the density of spins for $D=1.997$. It is possible to notice that the second peak does not scale with the lattice size and that the maximum of the second peak always falls around $T=0.80$.  
To understand which transition is occurring for each peak in the specific heat, we analyzed the density of spins $\pm1$ present in the lattice. It starts in $1$ and drops drastically around the pseudo-critical temperature, filling less than $10\%$ of the lattice sites at the first maximum in specific heat. The density of non-magnetic spins, $n_0=1-n_{\pm 1}$, goes in the opposite direction, drastically increasing in the region close to the pseudo-critical temperature. In fact, this increase in the number of spins $0$ after the pseudo-critical temperature manifests itself in a very moderate way for values of $D>1$, and gradually increases while increasing the crystal field.

This anomaly of the specific heat can be related to the presence of vacancies, or the Schottky defect, in a crystalline lattice \cite{kittel2005,w1976solid}, if the non-magnetic spins are interpreted as vacancies. The peak that presents finite-size effects, in general, is related to the order-disorder transition of the system. However, it can be seen in Figs. \ref{fig:cv-schot1}b and \ref{fig:cv-schot3}b, that after the transition, the number of spins $\pm 1$ increases again, which originates the second peak and does not have a scaling behavior. It is expected that at $T\to\infty$, the number of non-magnetic spins goes to $1/3$ while the number of $\pm 1$ spins goes to $2/3$. It can also be noted that the crossing of the $\pm 1$ spin number curves occurs at approximately the critical temperature.

A previous work with the Blume-Capel model\cite{Kwak2015wang}, using the Wang-Landau method, presented similar results. However, the authors found the same effect only for lattice sizes greater than $L=16$ and for crystal field $D \geqslant 1.990$, whereas here, even for the smallest simulated lattice size ($L=8$) the anomaly is already visible and for a smaller value of the crystal field $D \geqslant 1.950$. 

\section{\label{sec:level5}Conclusions}\label{sec:conclusions}

In this work, we have constructed the temperature-crystal field phase diagram of the Baxter-Wu model using entropic sampling simulations. 
The simulations were performed with a joint density of states, $g(E_1,E_2)$, so that with just one simulation run one calculates thermodynamic quantities for any temperature and crystal field values. To determine the point at which we have only discontinuous transitions, we observe the evolution of the maximum of the specific heat as a function of the crystal field. A peak is expected to appear for a certain value of the crystal field, from which the region of discontinuous transitions begins. This point was estimated as $D_{pc}=1.68288(62)$. The corresponding temperature at this point was determined as $ T_{pc} = 0.98030(10) $. These values are quite different from those found in\cite{Costa2004} $D_c=1.3089$ and $T_c=1.2225$, and\cite{Dias2017}, $D_c=0.890254$ and $T_c=1.1690$.
Finally, we also have observed a second peak in specific heat for crystal field values larger than $1.900$. This second peak does not scale with the size of the system, and appears to be due to the abrupt increase in the number of non-magnetic spins $s_i=0$, which can be associated with the Schottky defect in a crystalline lattice.

\section{\label{sec:level6}Acknowledgments}
We acknowledge the computer resources provided by LCC-UFG and IF-UFMT. L. S. F. acknowledges the support by CAPES, and L. N. Jorge acknowledges the support by FAPEG.

\bibliography{referencias.bib}

\end{document}